# Q-VR: System-Level Design for Future Mobile Collaborative Virtual Reality

Chenhao Xie
Pacific Northwest National Laboratory
USA

Xie Li
University of Sydney
Australia

Yang Hu
University of Texas at Dallas
USA

Huwan Peng
University of Washington
USA

Michael Taylor
University of Washington
USA

Shuaiwen Leon Song
University of Sydney
Australia

## ABSTRACT

High Quality Mobile Virtual Reality (VR) is what the incoming graphics technology era demands: users around the world, regardless of their hardware and network conditions, can all enjoy the immersive virtual experience. However, the state-of-the-art software-based mobile VR designs cannot fully satisfy the realtime performance requirements due to the highly interactive nature of user's actions and complex environmental constraints during VR execution. Inspired by the unique human visual system effects and the strong correlation between VR motion features and realtime hardware-level information, we propose *Q-VR*, a novel dynamic collaborative rendering solution via software-hardware co-design for enabling future low-latency high-quality mobile VR. At software-level, Q-VR provides flexible high-level tuning interface to reduce network latency while maintaining user perception. At hardware-level, Q-VR accommodates a wide spectrum of hardware and network conditions across users by effectively leveraging the computing capability of the increasingly powerful VR hardware. Extensive evaluation on real-world games demonstrates that Q-VR can achieve an average end-to-end performance speedup of **3.4x** (up to **6.7x**) over the traditional local rendering design in commercial VR devices, and a **4.1x** frame rate improvement over the state-of-the-art static collaborative rendering.

**ACM Reference Format:**
Chenhao Xie, Xie Li, Yang Hu, Huwan Peng, Michael Taylor, and Shuaiwen Leon Song. 2021. Q-VR: System-Level Design for Future Mobile Collaborative Virtual Reality . In *Proceedings of the 26th ACM International Conference on Architectural Support for Programming Languages and Operating Systems (ASPLOS '21), April 19–23, 2021, Virtual, USA.* ACM, New York, NY, USA, 13 pages. https://doi.org/10.1145/3445814.3446715

## 1 INTRODUCTION

Since the release of the movie *Ready Player One*, consumers have been longing for a commercial product that one day can levitate them to a fantasy alternate dimension: a truly immersive experience without mobility restriction and periodical motion anomalies. In other words, users require exceptional visual quality from an *untethered* mobile-rendered head-mounted displays (HMDs) that is equivalent to what high-end tethered VR systems (e.g., Oculus Rift [44] and HTC Vive [22]) provide. Although the current mobile hardware's processing capability has been significantly improved [3, 48], they still cannot fully process heavy VR workloads under the stringent runtime latency constraints. With the development of high performance server technology, server-based realtime rendering of Computer Graphics (CG) has been introduced by several major cloud vendors such as Nvidia GeForce Now [41] and Google Cloud for Game[17]. However, under the current network conditions, remote servers alone cannot provide realtime low-latency high-quality VR rendering due to the dominating communication latency. Thus, neither local-only rendering nor remote-only rendering can satisfy the latency requirements for high-quality mobile VR: *there is a clear mismatch between hardware's raw computing power and desired rendering complexity*.

To address the latency and bandwidth challenges of today's dominant mobile rendering models, it seems reasonable to utilize mobile VR hardware's computing power to handle part of the rendering workload near the display HMD to trade off for reduced network communication, while letting the remote system handle the rest of the workload. But how to design such VR systems to reach the latency and perception objectives is still an open problem. Recent studies [7, 31, 35–37] proposed a static collaborative software framework that renders the foreground interactive objects locally while offloading the background environment to the remote server, based on the observation that interactive objects are often more lightweight than the background environment. However, after a thorough qualitative investigating into the current mobile VR's architecture-level rendering pipeline and a quantitative latency bottleneck analysis, we observe that this naive rendering scheme faces several challenges.

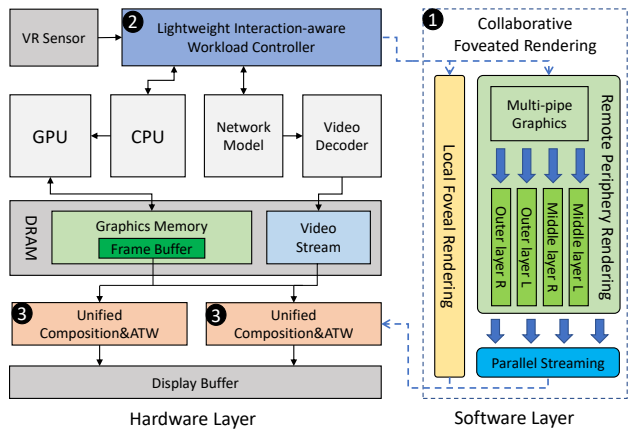

**Figure 1: Processing diagram of our software-hardware co-designed Q-VR.**

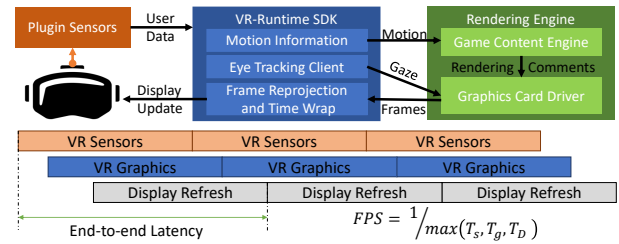

**Figure 2: An example of a modern VR graphics pipeline.**

First, interactive objects have to be narrowly defined by programmers on each hardware platform to satisfy the "worst case" scenario during VR application development which significantly limits the design possibilities for high-quality interactive VR environments and burdens programmers to accommodate all the realtime constraints during development. It is labor intensive and impractical. Second, it cannot fundamentally reduce the communication latency because the remote rendering workload remains unreduced. Third, it loses the flexibility to dynamically maintain the balance of local-remote rendering latency under realtime uncertainties: unpredictable user inputs (e.g., interaction, movements, etc.) and environment (e.g., hardware and network) changes. Finally, it suffers from high composition overhead by requiring more complex collision detection and embedding methods [7, 31], directly contributing to resource contention on mobile GPU(s).

In this paper, we propose a novel software-hardware co-design solution, named *Q-VR*, for enabling low-latency high-quality collaborative mobile VR rendering by effectively leveraging the processing capability of both local and remote rendering hardware. Fig.1 illustrates the processing diagram of our Q-VR. At the software-layer, we propose a vision-perception inspired collaborative rendering design ❶ for Q-VR to provide flexible tuning interface and programming model for enabling network latency reduction while maintaining user perception. The basic idea is that different acuity level requirements of human visual system naturally generate a new workload partitioning mechanism for collaborative VR rendering (Section 3). We leverage and extend this "foveation effect" [20, 51, 58–60] in Q-VR's software design to transform this complex global collaborative rendering problem into a workable framework. At the hardware-level, we design two novel architecture components, *Lightweight Interaction-Aware Workload Controller* (LIWC)❷ and *Unified Composition and ATW* (UCA)❸, to seamlessly interface with Q-VR's software-layer for achieving two general objectives: (1) quickly reaching the local-remote latency balance for each frame for the optimal rendering efficiency; and (2) forming a low-latency collaborative rendering pipeline for reducing realtime resource contention and improving architecture-level parallelism. These hardware designs are based on two key insights: there is a strong correlation among motion, scene complexity and hardware-level intermediate data (Section 4.1); and there is an algorithmic-level similarity between VR composition and reprojection (Section 4.2). To summarize, this paper makes the following contributions:

- We design the first software-hardware co-designed collaborative rendering architecture to tackle the mismatch between VR hardware processing capability and desired rendering complexity from a cross-layer systematic perspective;
- We identify the fundamental limitations of the state-off-the-art collaborative rendering design and quantify the major bottleneck factors via detailed workload characterization and VR execution pipeline analysis;
- By leveraging the foveation features of human visual system, we explore the software-level flexibility to reduce the network limitation via a fine-grained dynamic tuning space for workload control while maintaining user perception;
- Based on our key observations on VR motion correlations and execution similarity, we design two novel hardware components to support software-layer interfacing and deeper pipeline-level optimizations;
- Extensive evaluation on real-world games demonstrates that Q-VR design can achieve an average end-to-end speedup of **3.4x** (up to **6.7x**) over the traditional local rendering design in today's commercial VR devices, and a **4.1x** frame rate improvement over the state-of-the-art static collaborative rendering solution.

## 2 BACKGROUND AND MOTIVATION
### 2.1 The State-of-the-Art Mobile VR Systems

Different from the traditional graphics applications, modern VR systems retrieve the real-time user information to present a pair of realities scenes in front of users' eyes. Fig. 2 shows an example of a typical modern VR graphics pipeline. The VR system first gathers the head-/eye-tracking data at the beginning of a frame through plugin motion and eye sensors which are typically executed on their own frequencies [13, 20, 53]. Then, it relies on the VR runtime to process user inputs and eye-tracking information, and the rendering engine to generate the pair of frames for both eyes. Before the pair of rendered frames displayed onto the *Head Mounted Display* (or *HMD*), a VR system processes asynchronously time wrap (ATW) to reproject the 2D image plane based on *lens distortion*[5, 57]. To create a perception that users are physically present in a non-physical world (i.e., the concept of immersion [14, 27, 42, 57]), the rendering task becomes very heavy: generating a pair of high-quality images along with sound and other stimuli catering an engrossing total environment.

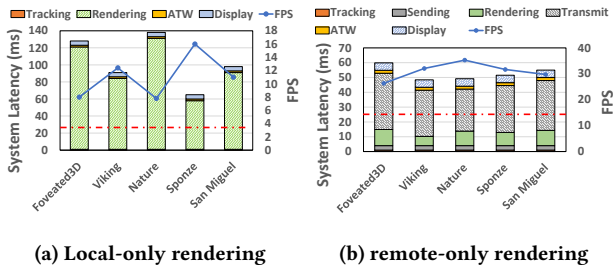

(a) Local-only rendering   (b) remote-only rendering

**Figure 3: System latency and FPS when running high-end VR applications on two current mobile VR system designs.**

Meanwhile, because the human vision system is very latency sensitive for close views, any noticeable performance degradation in VR real-time can cause motion anomalies such as judder, sickness and disorientation [7, 31]. To achieve robust real-time user experience, commercial VR applications are required to meet several performance requirements, e.g., the end-to-end latency (i.e., *Motion-to-Photon* latency or **MTP**) < 25 ms and frame rate > 90 Hz [27] (about 11ms) as Fig.2 demonstrates. In order to deliver high image quality simultaneously with low system latency, high quality VR applications are typically designed on a tethered setup (e.g., HTC-Vive Pro [22] and Oculus Rift [44]). The tethered setup connects the VR HMD with a high-end rendering engine (e.g., standalone GPU or GPUs) to provide desired rendering performance. However, bounded by the connection cable between VR HMD and render, tethered VR systems significantly limit users' mobility which is one of the core requirements of immersive user experience. With the advancement of mobile device design and System-on-Chip (SoC) performance, we have observed a trend of design focus shift from low-mobility rendering to a mobile-centric local rendering design, e.g., Google Daydream[18], Oculus Quest [44], Gear VR [50]. However, these rendering schemes cannot effectively support low-latency high-quality VR rendering tasks due to the wimpy mobile hardware's raw processing power compared to their tethered counterparts. As a result, the state-of-the-art mobile VR designs are limited to delivering VR videos instead of enabling real-time interactive VR graphics [32, 36].

## 2.2 Current Rendering Schemes for Mobile VR

With the development of wireless technology, the concept of cloud-based real-time rendering of Computer Graphics (CG) is being introduced by major cloud service vendors [2, 17, 41]. It opens up opportunities to stream VR games or other virtual contents from cloud servers to enable possible high-quality VR scene rendering on high-performance computing clusters [64]. There are two main rendering schemes proposed to support next-generation mobile VR rendering:

**(I) Remote Rendering.** A straightforward approach to overcome the performance limitation of mobile systems is to offload the compute-intensive rendering tasks to a powerful server or remote high-end GPUs by leveraging the cloud-based real-time rendering technologies. However, under the current network condition, the naive cloud VR design via streaming is infeasible to provide real-time high quality VR rendering due to the requirements of high resolution and low end-to-end latency. Previous work [13] suggests to leverage compression techniques to reduce the transmit latency. However, even with the highly effective compression strategies with parallel decoding, such approach cannot meet the performance requirements of high-quality VR applications[31].

Fig.3 shows the breakdown of the *end-to-end latency* (i.e., from tracking to display) for executing several high-quality VR applications under two commercial mobile VR designs: *local-only rendering* and *remote-only rendering*. The detailed experimental setup is discussed in Sec-2.3. The blue lines represent the frame rate (FPS) achieved on the VR HMD while the red dash lines illustrate mobile VR system latency restriction (i.e., the commercial standard of 25ms). The figure shows that the raw processing power of the integrated GPU is the key bottleneck for local-only rendering, while the transmission latency in remote-only rendering contributes to approximately 63% of the overall system latency.

Although the VR vendors today employ frame re-projection technologies such as Asynchronous TimeWarp (ATW) [5] to artificially fill in dropped frames, they cannot fundamentally reduce the MTP latency and increase the FPS due to little consideration of realtime inputs such as users' movements and interaction. Thus, improving the overall system performance is still one of the highest design priorities for future mobile VR systems.

**(II) State-of-the-Art: Static Collaborative VR Rendering.** Recent works [7, 31, 35, 37] have proposed a collaborative rendering scheme which applies mobile VR hardware's computing power to handle a portion of the **time-critical** rendering workload near the HMD display while letting the remote system handle the rest. Specifically, the fundamental principle of this collaborative scheme is based on the observation that the pre-defined interactive objects are often more lightweight than the background environment, suggesting to render the foreground interactive objects locally while offloading the background environment to the remote server. To further hide the network latency and improve bandwidth utilization, they also enable pre-rendering and prefetching for the background environment. However, this general scheme ignores several key factors, including (1) different mobile VR hardware's realtime processing capability, (2) ever-changing rendering workload due to realtime user inputs, (3) different network conditions available to users. These factors result in significant performance, programmbility and portability challenges for low-latency high-quality mobile VR design. We will discuss this in details next.

## 2.3 Analysis on Collaborative Rendering

**Rendering Execution Pipeline Analysis.** We first qualitatively analyze the general collaborative rendering and its limitations from the perspective of execution pipeline. Fig.4(top) describes a general collaborative rendering execution pipeline based on today's mobile VR design prototypes[7, 31, 35, 37]. A collaborative VR rendering workload can be interpreted as several functional kernels launched on to multiple accelerators [24] (with the same color in Fig. 4), each of which is responsible for executing a set of tasks. Specifically, for every frame, CPU utilizes VR input signal to process the VR application logic (CL). After that, it setups the local rendering tasks and issues remote frame fetching to the network (LS). Then the frame generation is split in to two parallel processes: the mobile GPU processes the interactive objects via local rendering (LR), while the network model offloads the background rendering to the remote server (RR). Then, the remote server returns the rendered background as encoded streaming network packets to be later decoded by the video processing unit and stored in the framebuffer (VD). When both the interactive objects and background image are ready, GPU composites them based on the depth information to generate the final frame (C). Since this output frame is still in 2D, GPU will further map it into 3D coordinates via ATW (lens distortion and reprojection) and deliver it to the HMD.

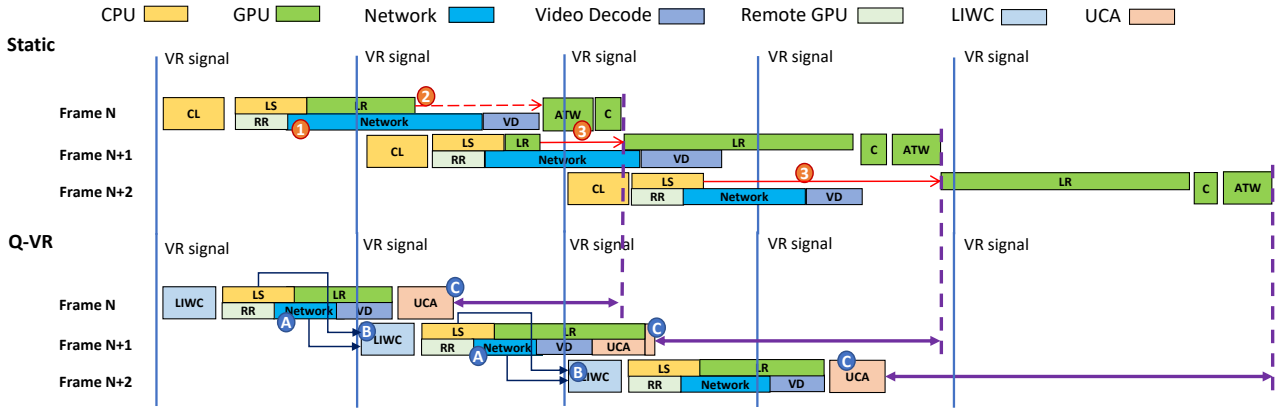

Figure 4: Execution pipeline of static collaborative rendering and our proposed Q-VR. Q-VR's software and hardware optimizations are reflected on the pipeline. Rendering tasks are conceptually mapped to different hardware components, among which LIWC and UCA are newly designed in this work. Intra-frame tasks may be overlapped in realtime (e.g., RR, network and VD) due to multi-accelerator parallelism. CL: software control logic; LS: local setup; LR: local rendering; C: composition; RR: remote rendering; VD: video decoding; LIWC: lightweight interaction-aware workload controller; UCA: unified composition and ATW.

Table 1: Performance of Static Collaborative VR rendering Across Different High-Quality VR Applications (90Hz)

| Apps | Resolution | #Triangles | Interactive Object | $f$ Range | Avg. $T_{local}$ | Min. $T_{local}$ | Max. $T_{local}$ | Back Size | $T_{remote}$ |
|---|---|---|---|---|---|---|---|---|---|
| Foveated3D[20] | 1920x2160 | 231K | 9 Chess | 16% - 52% | 43 ms | 18 ms | 75ms | 646KB | 38ms |
| Viking[56] | 1920x2160 | 2.8M | 1 Carriage | 10% - 13% | 13ms | 12ms | 16ms | 530KB | 31ms |
| Nature[55] | 1920x2160 | 1.4M | 1 Tree | 10% - 24% | 16ms | 12ms | 26ms | 482KB | 28ms |
| Sponze[42] | 1920x2160 | 282K | Lion Shield | 0.1% - 20% | 5.8ms | 0.5 ms | 12 ms | 537KB | 31ms |
| San Miguel[42] | 1920x2160 | 4.2M | 4 Chairs, 1 Table | 6% - 15% | 11 ms | 5.4 ms | 14 ms | 572KB | 33ms |

To achieve the highest rendering performance, both software-level parallelism (between different kernels) and hardware-level parallelism (between different hardware resources) need to be well coordinated. We identify **two general insights** for forming a low-latency collaborative rendering pipeline. (1) Within each frame, the local and remote rendering need to *reach a balance point* to achieve the highest resource utilization. The significant slowdown from either component will result in unsatisfactory execution and causing motion anomalies and low frame rate. For example, Fig4-❷ is caused by misestimating hardware's realtime processing capability and the changing workload during the execution. (2) Across frames, eliminating realtime GPU resource contention from different essential tasks can significantly improve framerate. As illustrated by Fig.4-❸, several essential tasks including local rendering, composition and ATW all compete for GPU resource. Any elongated occupation of GPU cores by composition and ATW can interrupt the normal local rendering process and cause bursts of frame rate drops. This phenomenon has been observed by previous studies [5, 32, 65].

**Challenges Facing Static Collaborated Rendering.** Now we investigate the design efficiency of the current static collaborative rendering. To provide quantitative analysis, we build our physical experimental platform for this evaluation. We execute several Windows OS based open source high-quality VR apps on a Gen 9 Intel mobile processor which is equipped with an Intel Core i7 CPU and a mobile GPU. We also calibrate the rendering performance of this local rendering platform against an Apple A10 Fusion SoC equipped by iPhone X[3] through executing a range of mobile VR apps. For remote rendering, a high-performance gaming system equipped with an NVIDIA Pascal GPU is used as the rendering engine. Additionally, Netcat [16] is applied for network communication and lossless H.264 protocol is leveraged for video compression.

Table 1 lists the tested high-quality VR applications and their performance characterization. This application is original designed for tethered VR devices and present photorealistic VR scenes. For each application, we first identify the draw batch comments for every object and then extract the foreground dynamic objects for local rendering and background for remote rendering as previous works[7, 31, 35, 37] suggest. The workload partition parameter,

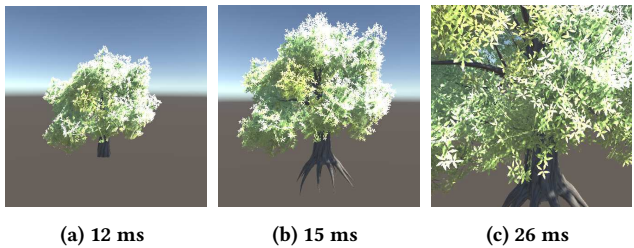

(a) 12 ms  (b) 15 ms  (c) 26 ms

**Figure 5: The realtime user inputs (e.g., interaction) directly affects latency to vary even within the same scene. The closer to the tree in Nature[55], the more details need to be rendered.**

$f$, represents the percentage of the normalized latency to render the interactive objects to the entire frame rendering time. We also collect the latencies for the local rendering ($T_{local}$), remote frame fetching ($T_{remote}$) which should smaller than 11 ms to satisfy 90Hz FPS. Since the remote rendering, network transmission and video codex can be streamed in parallel [31, 34], we only count the highest latency portion from the remote side which is the network transmission in our case. From Table 1, we have identified **two major challenges** for static collaboration:

**Challenge I: Design Inflexibility and Poor Programmability.** The state-of-the-art design is a "one-fit-for-all" solution: it assumes the processing of the pre-defined interactive objects will always meet VR's realtime latency requirements. However, the VR scene complexity and animation of interactive objects are often random and determined by users' actions at realtime which may cause significant workload change from frame to frame. Fig.5 and Table 1 demonstrate that the rendering latency for a single interactive object (the tree in the Nature app) can change from 12ms to 26ms (i.e., 10% - 24% rendering workloads) depending on how users interact with the tree, and the maximum $T_{local}$ of all benchmarks exceed the fps requirement (11$ms$ or 90HZ). As a result, in this static collaborative design, programmers are burdened to accommodate all the realtime constraints and reduce the interactive concepts in their developing to avoid VR latency issues, which is extremely difficult, labor intensive and impractical. Additionally, this design loses the flexibility to control runtime kernel execution (e.g., in Fig.4-②, transmission latency is long) to help local and remote rendering reach a balance point for optimal rendering and resource utilization.

**Challenge II: Costly Remote Data Transmission.** Table 1 also shows that the static design incurs high network latency (about 30ms in WIFI) to download the compressed background image, which significantly increases the end-to-end latency (demonstrated in Fig.4-①). Under this design, not only the rendered frames, but also the depth maps of the VR scenes need to be sent back for composition [7, 31, 35, 37]. Although the static collaborated rendering enables caching and prefetching techniques [7, 31] attempting to hide the network latency under some circumstances, they encounter large storage overhead. Meanwhile, to prefetch the background in time, mobile VR systems need to predict random users' motion inputs more than 30 ms ahead (about 3 frames) which may significantly reduce the prediction accuracy. Furthermore, failing to predict users' behaviors will trigger even higher end-to-end VR latency, resulting in motion sickness from the position mismatch between the interactive objects and background. [29, 46].

To tackle these challenges above, we propose a novel software-hardware co-design solution for low-latency high-quality collaborative VR rendering, named *Q-VR*. Its general pipeline is shown in Fig.4 (bottom). Based on the insights from this subsection, Q-VR has the following high-level designing objectives: (a) reducing $T_{remote}$ to weaken the impact of remote rendering and network latency; (b) dynamically balancing local and remote rendering based on realtime constraints (e.g., hardware, network and user inputs) for optimal resource utilization and rendering efficiency; and (c) eliminating realtime hardware contention on the execution pipeline to improve FPS. We breakdown Q-VR's design into a new software framework (Sec.3) and novel hardware supports (Sec.4).

## 3 EXPLORING SOFTWARE-LEVEL FLEXIBILITY FOR COLLABORATIVE VR RENDERING

In this section, we propose a *vision-perception inspired software layer design* for our Q-VR to provide a flexible interface for enabling $T_{remote}$ reduction while maintaining user perception. It also provides high-level support for the fine-grained dynamic rendering tuning capability enabled by our hardware design optimizations (Sec.4) which effectively accommodates rendering workload variation across frames and help reach local-remote latency balancing.

Instead of predefining the workload partition during VR application development, we extend the concept of *foveated rendering* [20, 28, 43, 60] to redesign the rendering workload for mobile VR systems. Previous research has documented how human visual acuity falls off from the centre (called *fovea*) to the *periphery*[49, 52]. Although human eyes can see a broad field of view (135 degrees vertically and 160 degrees horizontally), only *5 degrees central fovea area* requires fine details. For the *periphery* areas, the acuity requirement falls off rapidly as eccentricity increases. Based on this feature, foveated rendering can reduce rendering workload via greatly reducing the image resolution in the periphery areas and is able to maintain user perception as long as foveated constraints are satisfied between layers [9, 20, 38, 47, 51].

**Basic idea.** The basic idea is that the varying spatial resolution requirements in the human visual system (e.g., fovea versus peripheral vision) naturally generate an efficient workload partitioning. We can leverage this to significantly reduce the transmitted data size on the network through adapting lower resolutions of video streaming for periphery area on the remote server, but also effectively render the *most critical* visual perception area locally with the highest resolution without any approximation.

### 3.1 Runtime-Aware Adaptive Foveal Sizing

Traditional foveated rendering decomposes the frame into three layers: (1) the *foveal layer* (has a radius of $e_1$) in the eye tracking center which is the most critical perception area with the highest resolution; (2) the *middle layer* (has a radius of $e_2$) which employs gradient resolution to smooth the acuity falling; and (3) the *outer layer* which renders the periphery area with low resolution for speedup. Many past user perception surveys [1, 20, 38] have demonstrated that foveated rendering determines the resolutions following a well selected MAR (*minimum angle of resolution*) model to achieve the *same perceptive visual quality* with non-foveated rendering.

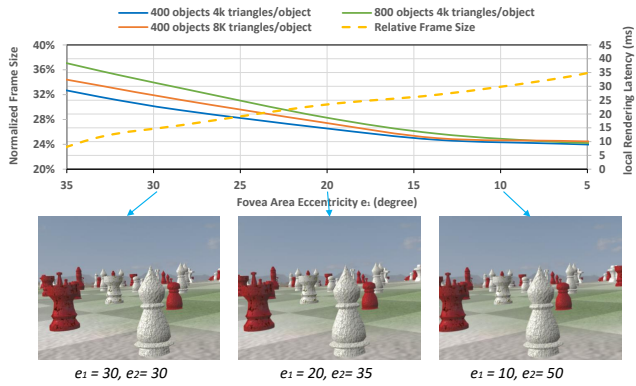

Figure 6: Average foveal layer rendering latency under the increasing eccentricity when running Foveated3D on Intel Gen9 mobile processor. When the eccentricity is ≤ 15 degrees, all types of scene complexities can be handled within the target latency requirements (≤ 11ms)

To estimate the local SoC's computing capability, we evaluate the rendering latency (end-to-end) according to foveal layer radius by executing Foveated3D app on a state-of-the-art Intel Gen 9 mobile processor and remote server collaboration setup (Sec.2.3). Here we reorganize the three layers into two: the local fovea rendering for the centre ($e_1$) and the remote periphery rendering for middle and outer layers ($*e_2$). We also adapt the second eccentricity ($*e_2$) and calculate the *\*Periphery Quality* via Eq.(1) to further reduce the communication overhead.

$$* e_2 = min\ (P_{Middle} + P_{Outer}) \\ * s_i = \frac{\omega_i}{\omega^*} = \frac{m * e_i + \omega_0}{\omega^*}, \quad i \in \{1, 2\} \quad (1)$$

where we directly employ the vision parameters (e.g., MAR slope $m$, fovea MAR $\omega_0$) from the previous user studies [1, 20, 38] to maintain user perception within the foveated constraints.

Fig.6 demonstrates that the local rendering performance highly depends on the size of the foveal layer. We observe that if the eccentricity is ≤ 15 degrees, all types of scene complexities in Foveated3D can be handled within the target latency requirements (≤ 11$ms$). This suggests that modern VR mobile SoCs are capable of dynamically rendering a range of workloads (or fovea sizes) with fine details and high resolution beyond the traditionally defined 5 degrees central fovea, determined by realtime constraints such as scene complexity, hardware capability, etc. This finding provides a flexible tuning knob for enabling dynamic workload control for Q-VR and helps further deprioritizing network latency and remote rendering.

Finally, we conduct an image quality survey following the evaluation principles from [20, 29] to evaluate the impact of our eccentricity selection method. Specifically, we take a user survey to 50 candidates to estimate the image quality effects after adapting our adaptive foveated rendering scheme. First, We apply different VR steam of images under a specific display resolution (e.g., 1920x2160) with different fovea areas (i.e., referring to the eccentricity from 40 degrees to 5 degrees) and their corresponding periphery resolutions. We then let the candidates focus on the center of the images and switch images based on the degrading central foveal eccentricity. Each image will be displayed for 5s. We then ask them if they experience any image quality difference and let them score each image during the survey. Similar to what is reflected in Fig.6 from different snapshots of the chessboard, participants observe no visible image quality difference between different eccentricity selections when the target MAR is satisfied which helps Q-VR maintain user perception.

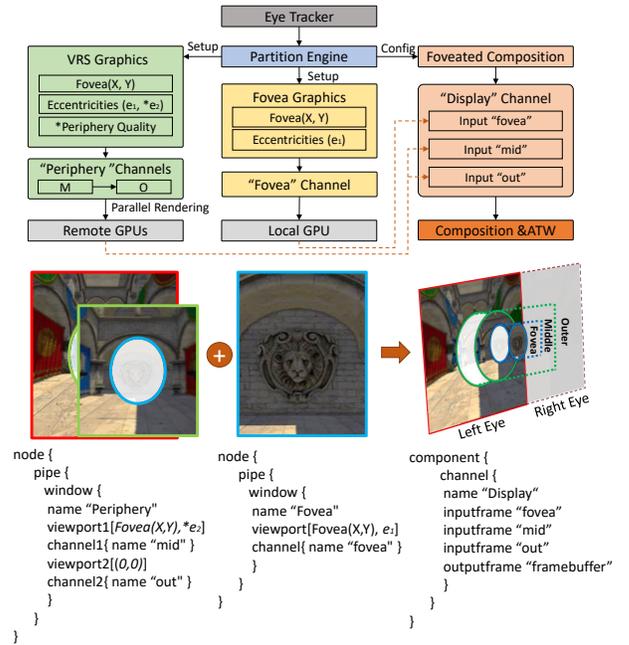

Figure 7: An example of software-level setup and configuration in our vision-perception inspired Q-VR, its programming model, and how it interfaces with hardware.

### 3.2 New Software Framework

We then introduce the software-layer support for enabling this fovea-ted-oriented collaborative rendering for future mobile VR. Different from the original foveated rendering focusing on image resolution approximation with pre-calculated eccentricities and resolutions, the design goals of the new software framework is to enable a dynamic partition by leveraging the key observation that the central fovea size depends on real-time hardware rendering capability. To achieve this, we created a new distributed rendering programming model supported by lower-level graphics libraries.

Fig.7 shows the overall software-layer design of our proposed Q-VR to support collaborative foveated rendering. First, we split the VR graphics into a local client version (the yellow boxes) and a remote server version (the green boxes) to process different visual layers in parallel. Instead of directly collecting the foveated rendering parameters such as the central fovea coordinate *foveat(X,Y)* and the partition *eccentricity* ($e_1,e_2$) from the eye tracker, we add a software tuning-knob for fine-grained fovea control and software interfaces to the graphics to acquire these parameters from our hardware partition engine, which is integrated into the workload controller described in Section 4.1. For the client version, we gather the $foveat(X,Y)$ and $e_1$ to setup the rendering viewports via VR SDK and the local rendering process remains as normal VR rendering for the two eyes in high resolution. For the server version, we extend the state-of-the-art parallel VR rendering pipeline [6, 11, 12, 43, 64] to setup multiple rendering channels for middle and outer layers with calculated eccentricity ($e_1,*e_2$).

Since Q-VR requires no additional composition on the remote server (supported by our UCA design in Sec.4.2), we use separated framebuffers to store the rendering results from the periphery layers. Each framebuffer has an adjustable size according to its corresponding layer's resolution or periphery quality. By doing this, the server only needs to send the lower quality middle and outer layers (under the fovated visual constraints though) back to the local client instead of the entire framebuffer with full resolution to reduce the transmitted data size. Using separated framebuffers, we apply parallel streaming to transmit data for middle and outer layers for each eye (Fig.7) and overlap the rendering and data transmit to further reduce the transmit latency. Finally, we performing the foveated composition to simply overlap the three layers' inputs. To erase the artificial effects generated by the resolution gradient between layers, the composition also processes multi-sample anti-alias (MSAA) on the edge [20] of layers. We discuss our novel hardware supports next.

## 4 HARDWARE SUPPORT FOR FINE-GRAINED RUNTIME CONTROLLING AND PIPELINE OPTIMIZATIONS

By proposing the software-layer design, we enable the possibility of realtime tuning rendering workload via adaptive foveal sizing. However, the actual eccentricity selection for each frame requires high fidelity and ideally should has minimized latency, which only software-based control mechanism cannot provide. As shown in Fig.4, to dynamically predict the proper fovea size, software control logic (CL) has to wait until the previous rendering completes which may delay more than one frame, e.g., Frame N+2's prediction is based on Frame N's rendering output. This not only causes low prediction accuracy but also may significantly extend the overall execution pipeline. This motivates us to explore hardware-level opportunities for deeper pipeline-level optimizations.

### 4.1 Lightweight Interaction-Aware Workload Controller (LIWC)

A straightforward method to dynamically select the best eccentricity would be statically and exhaustively profiling various parameters (e.g., hardware and network conditions, fps and MTP, user actions, etc) for each frame's eccentricity set ($e_1, e_2$) in a large sampling space, and build a model to predict $e_1$ for each frame. In reality, however, correctly predicting such mapping is very difficult because there is a large number of samples required even for a single scene [7] and is not portable to the other VR applications. Recent approaches [40, 61] have used deep learning models to train certain dynamic relationships but they are too power hungry to be integrated in mobile VR. Thus, we propose a *lightweight* design that can largely describe scene complexity change and help dynamically build a strong mapping between environment conditions and $e_1$.

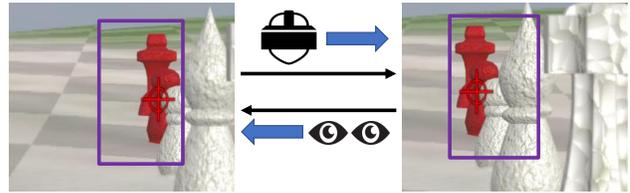

**Figure 8: The head motions and fovea tracking can help determine the scene complexity change trend across frames.**

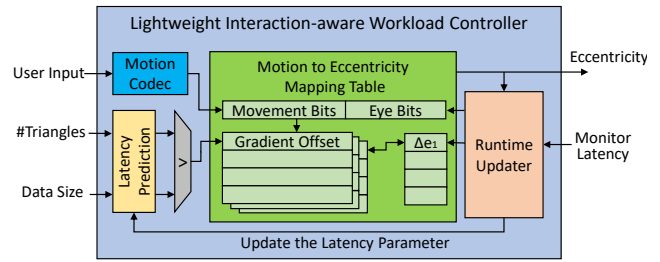

**Figure 9: Architecture diagram of our proposed LIWC.**

**Key Design Insights.** To build such mapping, we leverage *two key observations*. (i) The scene complexity change for the local foveated rendering across continuous frames is highly related to user's head and eye motions. Fig.8 shows an example: the center focus moves relatively with user's head and eyes to the left and right which changes the rendering workload in the fovea area (the purple box) accordingly. This indicates that it is possible to use this built-up interaction experience to correlate change trend for scene complexity with fovea area movements. (ii) The local rendering latency is sensitive to the scene complexity and realtime hardware processing capability (e.g., can be estimated by triangles[1]) while the remote latency is dominated by the resolution and network bandwidth. To respond to the environmental changes as soon as possible with minimal latency impact on the overall execution pipeline, we can predict the local and remote latency by *directly leveraging the intermediate hardware information*.

**Architecture design.** Based on these two insights, we propose a *lightweight Interaction-Aware Workload Controller* (named **LIWC**) shown in Fig.9, to determine the best balanced eccentricity which is indexed by user's inputs and runtime latency. It includes four major components: (1) an SRAM to store the motion-to-eccentricity mapping table which records the *latency gradient offset* for all pairs of motion information and eccentricity; (2) a latency predictor to predict the current latency for the local and remote rendering; (3) a motion codec to translate the motion information into table entry addresses; and (4) a runtime updater to update the mapping table and latency prediction parameters.

---
[1]Triangles are the basic intermediate units in computer graphics pipeline for creating more complex models and objects.

As a single accelerator separated from CPU and GPU, LIWC can bypass the CPU to directly monitor *the number of triangles* during the rendering setup process for assessing the local rendering latency, and to monitor the network's ACK packets for assessing the remote latencies. Leveraging these two hardware-level intermediate data, the local and remote latencies are estimated based on a lightweight performance model, described as Eq.(2). As Fig.4-Ⓑ illustrates, LIWC design avoids the overheads that the software approaches introduce, e.g., waiting for the rendering to complete, in-out memory activities, and kernel issuing.

$$T_{local} = \frac{\#Triangles * \%fovea}{P(GPU_m)}, \quad T_{remote} = \frac{DataSize\,(M+O)}{Throughput} \quad (2)$$

To gather user's inputs, LIWC indexes the motion information with the changes of user motion between two frames (i.e., 6 bits for degrees of freedom changes on HMD and 4 bits for the fovea center movement) through motion codec. This is to strictly control the parameter space size for both motion and eccentricity coordinates, since the motion information and the eccentricity mapping have an infinite parameter space when the problem scales up. Similarly, LIWC also indexes the eccentricity with a set of integer delta tags (-5 to 5 degrees) for each motion entry.

During the eccentricity selection, LIWC looks up the table entry with the closest latency gradient offset from the motion to eccentricity mapping table based on the motion index and the estimated latency difference between the local and remote rendering. After taking the selected delta eccentricity, the runtime updater monitors the realtime measured latency and the change of FPS to online update the latency gradient offset with a reward function ($gradient = (1 - \alpha) * gradient' + \alpha * \Delta latency$, where $\alpha$ represents the reward parameter and $gradient'$ represents the original latency gradient). It also updates the network throughput and GPU performance for further latency prediction. The table and parameters update phase will be executed in parallel with composition and display for minimizing the overall rendering latency.

### 4.2 Unified Composition and ATW Unit

As discussed Fig.4-③ in Sec.2.3, resource contention between rendering and composition/ATW on realtime GPU resources across frames result in delaying critical rendering process and cause significant FPS drops. One challenge is how to conduct parallel rendering of the complex scenes on Q-VR with efficient composition and ATW execution to form a low-latency collaborative rendering pipeline.

**Key Design insight: Algorithmic-level Similarity.** Fig.10-(top) shows the traditional sequential execution of composition and ATW. To smooth the resolution gaps between layers, the original foveated rendering performs anti-aliasing by combining the pixel colors from the rendered frames of the two layers. It calculates the average pixel color using Eq.(3)-(left) and then ATW fetches the composited frame from the framebuffer in GPU memory as a texture. After this, the frame is mapped into a sphere plane based on HMD lens distortion (2D to 3D) and the coordinate reprojection map (update to the latest motion position). During ATW, the plane frame is cut into small tiles (32x32) for SIMD execution and then fed into a specialized texture filter for bilinear filtering (Eq.3-(right)). From the two equations in Eq.(3), we observe a **key design insight** that if ATW is first processed for multiple vision layers then fed into composition (i.e., reordering in Eq.4-right), these two filtering phases can be *combined to a unified filtering process* which only samples the input once. In computer graphics, the unified filtering process can be operated as trilinear filtering.

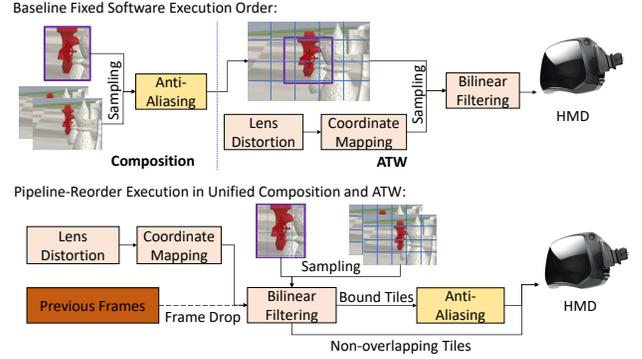

**Figure 10: Comparison between baseline sequential execution and Unified Composition and ATW (UCA).**

The advantages of using a unified process include: (1) it bypasses CPU and avoids the software overhead between kernels; (2) it breaks the fixed software execution sequence so that the ATW can start processing the non-overlapping tiles (e.g., tiles require no composition) earlier; and (3) it can be executed in parallel with GPU for better parallelism.

$$\text{Composition: } X = \frac{1}{M} \sum_{i}^{M} S_i, \quad \text{ATW: } Y = \frac{1}{N} \sum_{i}^{N} w_i * X_i \quad (3)$$

$$Y = \frac{1}{N} \sum_{i}^{N} w_i * \left(\frac{1}{2} \sum_{j}^{2} S_{ij}\right) = \frac{1}{MN} \sum_{j}^{M} \sum_{i}^{N} w_i * S_{ij} \quad (4)$$

**Architecture design.** Due to the algorithm-level similarity between ATW and composition, we propose to use a single Unified Composition and ATW Process (UCA) to replace the two independent computation paths by combining ATW with the unique fovea composition, and asynchronously executing them across frame tiles prior to the rendering completion (Fig.4-Ⓒ). Fig.10-(bottom) shows the execution pipeline of the proposed architecture. Unlike the original VR pipeline which separates the frame composition and re-projection, the new unified kernel reorders the *filtering stage* (i.e., first processing ATW for multiple vision layers then fed into composition) and combines them into a Trilinear filter with the same inputs of original foveated composition. The UCA can also leverage the previous frame layers to artificially reconstruct the updated frame with a new position as what the original ATW outputs. This helps fill in dropped frames to avoid coordination errors between two layers.

At hardware-level, we implement the UCA as a separate hardware unit on SoC to eliminate possible large and burst latency scenarios caused by GPU resource contention. We reused some of the logic units from the state-of-the-art ATW design[5, 32, 65] for lens distortion translation, coordination mapping and filtering. The UCA Unit mainly consists of two microarchitecture components: 4 MULs for lens distortion and 8 SIMD4 FPUs for coordination mapping and filtering. Fig.11 shows the architecture diagram of the proposed UCA. By monitoring the video stream and the framebuffer signals, UCA can detect if the data is ready in the DRAM. When the data is ready, UCA acquires the motion information from the HMD sensors and processes lens distortion and coordinates mapping as those in the normal ATW procedure. Then, it checks if the block belongs to the border of the two layers. For the border tiles, UCA processes an single trilinear filtering as eq.4 and sends the results back to the framebuffer. For the non-overlapping tiles, UCA directly processes them via bilinear filtering to generate the final pixel color.

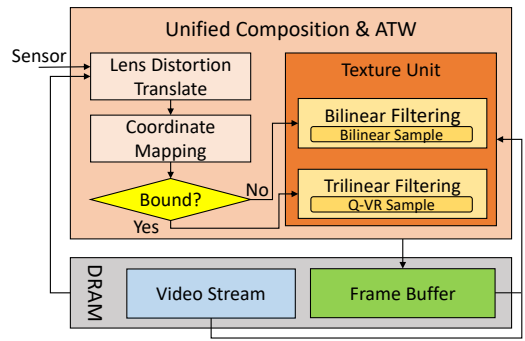

Figure 11: Architecture diagram of UCA.

Table 2: BASELINE CONFIGURATION

| Mobile VR System | |
|---|---|
| GPU frequency | 500 MHz |
| Number of Unified Shaders | 8 |
| Shader Configuration | 8 SIMD4-scale ALUs |
| | 16KB Unified L1 cache |
| | 1 texture unit |
| Texture Filtering Configuration | 4x Anisotropic Filtering |
| Raster Engine | 16x16 tiled rasterization |
| L2 Cache | 256 KB in total, 8-ways |
| DRAM Bandwidth | 16 bytes/cycle |
| | 8 channel |
| Unified Composition and ATW Unit | |
| Frequency | 500 MHz |
| Count | 2 |
| Remote GPU | |
| GPU Configuration | Multi GPU system as [64] |
| Network Throughput (Download Speed) | |
| Wi-Fi | 200 Mbps |
| 4G LTE | 100 Mbps |
| Early 5G | 500 Mbps |

### 4.3 Design Overhead Analysis

We use McPat to evaluate the area and power overhead of our proposed architecture. For LIWC, the SRAM table dominates its area and power cost. Due to our cost-effective design, its memory depth can be as small as $2^{15} = 32768$. We use a 16 bit half-precision floating-point number to represent the latency gradient offset, and the size of the table is estimated as approximately 64KB which has 0.66 $mm^2$ area overhead and maximum 25 mW power overhead under the default 500Mhz core frequency and 45nm technology by McPat[33]. For UCA, we reference previous works[32, 65] to map the logic units to hardware architecture. The McPAT results show that a single UCA occupies an area of $1.6mm^2$ and consumes 94mW runtime power at 500 MHz. For the latency overhead, since we formulate our eccentricity selection into a lightweight table lookup, the computation in the latency prediction and parameter updating are quite simple. We estimate the latency per frame can be as low as nanoseconds level. Thus, LIWC's latency overhead can be completely hidden. Additionally, we implement UCA as a texture mapping unit on a cycle-level mobile GPU simulator. Under the default configuration (Sec.5), the latency to process one 32x32 pixels block in UCA can be as low as 532 cycles. With 2 UCAs operating at 500 Mhz, we are able to achieve sufficient performance for realtime VR.

Table 3: BENCHMARKS

| Names | Library | Resolution | #Batches |
|---|---|---|---|
| Doom3-H | OpenGL[45] | 1920x2160 | 382 |
| Doom3-L | OpenGL | 1280x1600 | 382 |
| HL2-H | DirectX[39] | 1920x2160 | 656 |
| HL2-L | DirectX | 1280x1600 | 656 |
| GRID | DirectX | 1920x2160 | 3680 |
| UT3 | DirectX | 1920x2160 | 1752 |
| Wolf | DirectX | 1920x2160 | 3394 |

## 5 EVALUATION METHODOLOGY

**Evaluation Environment.** To model the proposed Q-VR software layer and hardware design, we use similar validation methods from the previous work[62–64] on a modified ATTILA-sim[4], a cycle-accurate rasterization-based GPU rendering simulator which covers a wide spectrum of modern graphics features. Specifically, for the rendering pipeline, we implement simultaneous multi-projection engine in ATTILA-sim to support two-eyes VR rendering and reconfigure it by referencing the ARM Mali-G76 [10], a state-of-the-art high-end mobile GPU. Following the design details from Section 3, we separately implement the client and server version of our Q-VR framework in ATTILA-sim by modifying the GPUDriver and the command processor. The added architecture blocks, including LIWC and UCA, are implemented as a rendering task dispatcher and a post-processor, respectively. They are also integrated into the rendering pipeline in ATTILA-sim. We also investigated other detailed hardware latencies (e.g., eye-tracking, screen display, etc) and integrate them into our model for an enhanced end-to-end simulation. For instance, since the eye-tracking latency is not in the critical path of the graphics pipeline (Section 7), we count 2ms to transmit the sensored data to the rendering engine and 5 ms to display the frame on HMD[13, 20] in the end-to-end latency.

For evaluating the network transmission latency, we leverage ffmpeg [15] to compress the output frames from the remote server and then use them to estimate network latency based on different downloading speeds. The network latency is calculated by dividing the network bandwidth with the compressed frame size. Furthermore, we insert white noises into our network channel with 20dB SNR (Signal-to-Noise Ratio) to better reflect reality. We validate our model against netcat [40] which is widely used in linux backends to build communication channels and found that our network model is able to reflect the real communication channels to a great extent. For the remote server side, we implement a future chiplet based multi-GPU design that can scale up to 8 MCM GPUs (similar to that in [64]) to enable high performance parallel rendering. Table 2 illustrates the simulation configuration and network throughput used in our evaluation. We choose 500 MHz and Wi-Fi as our default GPU core frequency and network condition, respectively.

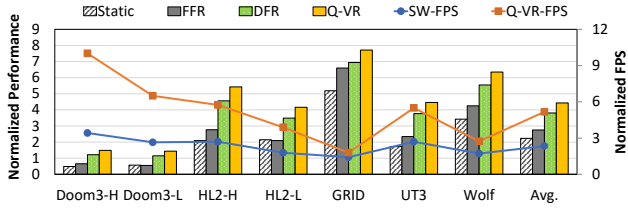

Figure 12: The normalized performance improvement from different designs under the default hardware and network. The results are normalized to the traditional local rendering design appeared in today's mobile VR devices.

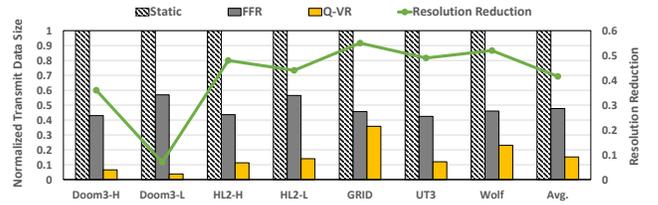

Figure 13: The normalized transmitted data size and resolution reduction from different designs under the default hardware and network. The results are normalized to the remote rendering design in commercial cloud servers.

**Benchmarks:** Table 3 lists a set of gaming benchmarks employed to evaluate Q-VR. This set includes We employ five well-known 3D games from ATTILA-sim, which are well compatible with our simulator, to evaluate Q-VR. The benchmarks set covers different rendering libraries and 3D gaming engines [39, 45]. Although the graphics API traces can be directly used for evaluation, we do adjust the entire frame resolution per eye to match the setting in our VR HMD. To better understand the effectiveness of Q-VR, two benchmarks (Doom3 and Half-Life 2) are rendered with both low and high resolutions (1920 × 2160, 1280 × 1600); while for the others (UT3, GRID and Wolf), 1920 × 2160 is adopted as the baseline resolution.

## 6 RESULTS AND ANALYSIS
### 6.1 Overall Performance Improvement

We first estimate the performance improvement of Q-VR by comparing it with several design choices under the default hardware and network condition: (i) *Baseline* – traditional local rendering in commercial VR device. (ii) *Static* – static collaborative VR rendering which leverages mobile GPU to render the interactive object from frame $n$ and prefetching the background of frame $n + 1$ from remote GPUs. We identify the interactive object in ATTILA-sim by comparing the depths of all rendering batches and find the closet one to viewports; (iii) *Fixed Foveated Rendering (FFR)* – collaborative foveated rendering with static eccentricity based on the classic MAR model (i.e., $e_1 = 5$, traditional fovea size), discussed in Section 3 ; (iv) *Dynamic Foveated Rendering (DFR)* – collaborative foveated rendering with only LIWC enabled; and (v) *Q-VR* – our proposed collaborative VR rendering.

**End-to-End System Latency.** Fig.12 shows the normalized speed-ups of different design choices over the Baseline case, the traditional local rendering in commercial VR device. we calculate the average end-to-end system latency from each design and normalize them to the pure local rendering case. From the figure, we make the following observations. First, naively partitioning the fovea and periphery area in *FFR* design is able to achieve approximately 1.75x and 52% performance improvement on average and up to 5.6x and 1.4x over the baseline and static, respectively. This is because even under the fixed fovea area, Q-VR design software framework is able to reduce a certain amount of data transmitted back from the server via resolution approximation. However, the speedup by FFR can be limited by the network latency. We observe that for most of the benchmarks, network latency is much higher than the local rendering latency under FFR design. In other words, the latency balance is not reached. Third, by leveraging our LIWC design, DFR is able to reach a more balanced state: it achieves an average of 1.1x speedup over FFR. Finally, by leveraging UCA to further extend the accelerator-level parallelism over FFR, Q-VR outperforms others and achieves an average of 3.4x speedup (up to 6.7x) over Baseline.

**Frame Rate.** We also compare the frame rate (FPS) achieved between a pure software design and our proposed software-hardware co-design. We build the pure software implementation of Q-VR by selecting eccentricity based on previous local and remote rendering latency instead of using the intermediate hardware data (e.g., #triangles and network condition) for predicting rendering and network latencies in LIWC even prior to rendering completion (Fig.4-Ⓑ). The two solid lines in Fig.12 shows the average FPS improvement of the pure software implementation (SW-FPS) and our Q-VR (Q-VR-FPS) over the Baseline. We calculate FPS as $FPS = min(1/T_{GPU}, 1/T_{network})$. The result demonstrates that Q-VR outperforms the static collaboration design and software implementation by 4.1 × and 2.8 ×, respectively. First, Q-VR achieves better latency balancing than the pure software design by leveraging the intermediate hardware data to fast and accurately predict the best eccentricity. Additionally, by detaching the ATW and composition processes from GPU core execution, Q-VR can increase GPU utilization for rendering and better exploit multi-accelerator level parallelism.

**Network Transmission.** Fig.13 shows the normalized transmitted data size and resolution reduction from different designs under the default hardware and network condition. The results are normalized to the remote rendering in commercial cloud server. From the figure, we observe that the static approach does not reduce the actual transmitted data size. Alternatively, it prefetches the backgrounds to hide the network latency. Compared to the static collaborative rendering, Q-VR achieves an average transmitted data reduction of 85% by runtime adopting optimal foveal sizes and reducing the periphery area resolutions. Regarding the overall resolution reduction, Q-VR achieves an average of 41% reduction over the original frame. We want to emphasize that the transmitted data size reduction does not only originate from resolution reduction; it also comes from correctly adjusting the central fovea workload on the local hardware based on different realtime constraints. For example, Q-VR reduces 96% transmitted data size for Doom3-L with 7% resolution reduction. Since Doom3-L is the lightest workload in our experiments, most of the rendering work is executed locally.

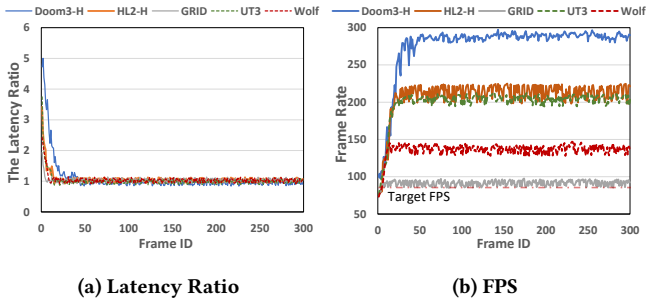

(a) Latency Ratio  (b) FPS

Figure 14: The Latency Ratios and FPS across 300 Frames.

Table 4: Best Eccentricity Under Different Configurations

| Freq. | Net. | Benchmarks | | | | | | |
|---|---|---|---|---|---|---|---|---|
| | | D3H | D3L | H2H | H2L | GD | NFS | WF |
| 500 MHz | Wi-Fi | 46.4 | 85.3 | 27.4 | 33.2 | 9.9 | 27.2 | 15.3 |
| | 4G LTE | 74.5 | 90 | 42.2 | 44.3 | 22.1 | 39.1 | 25.7 |
| | Early 5G | 22.4 | 45.2 | 11.3 | 14.3 | 5 | 10.9 | 8.6 |
| 400 MHz | Wi-Fi | 34.5 | 77.3 | 23.1 | 26.1 | 7.8 | 22.5 | 13.2 |
| | 4G LTE | 64.3 | 90 | 34.5 | 39.2 | 15.5 | 32.4 | 18.5 |
| | Early 5G | 15.3 | 30.2 | 7.8 | 11.5 | 5 | 7.4 | 6.1 |
| 300 MHz | Wi-Fi | 27.5 | 65.4 | 16.4 | 24.5 | 6.5 | 14.3 | 11.3 |
| | 4G LTE | 43.2 | 90 | 30.2 | 35.1 | 12.4 | 27.2 | 16.4 |
| | Early 5G | 13.1 | 27.1 | 6.9 | 8.3 | 5 | 6.1 | 5 |

### 6.2 Local and Remote Latency Balancing in Q-VR

To evaluate if our Q-VR can help the rendering pipeline quickly reach the balanced local-remote latency state under different user inputs and environment constraints, we calculate the latency ratio ($T_{remote}/T_{local}$) for each frame during a game execution, as shown in Fig.14-(a). We initiate Q-VR with $e_1 = 5$ under the default hardware and network condition. From the figure, we observe that the latency ratios are quite high during the first several frames. This is because relatively small eccentricity makes the local hardware to render quite fast while the network latency becomes the primary bottleneck which causes local-remote latency imbalance. The figure also demonstrates that Q-VR can gradually locate the balanced eccentricity to reach the best rendering efficiency after a very short period of time. Finally, Fig.14-(b) proves it is able to maintain very high FPS for all benchmarks which satisfy the high-quality VR requirement (>90Hz).

### 6.3 Sensitivity Study

**Eccentricity Selection Under Different Configurations.** Table 4 shows the average eccentricity (i.e., $e_1$ radius value) selected by Q-VR across different applications and hardware/network conditions. We started recording the eccentricity for each frame after Q-VR reaches a steady state and then calculate their average. Note that scene complexity can dynamically change from frame to frame.

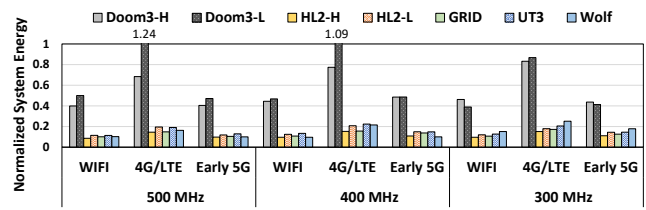

Figure 15: The normalized energy efficiency of Q-VR under different hardware and network conditions.

From the table, we observe that under different configurations the average eccentricity can be quite different. For example, under the default GPU frequency and Wi-Fi, Doom3-L has a much bigger $e_1$ than GRID. This is because GRID has more complex scenes than Doom3-L and requires longer rendering time. Thus Q-VR keeps the eccentricity small and giving more workload to remote GPUs to balance local-remote latency for the best rendering performance. Similar situation occurs when increasing network throughput or reducing GPU frequency. Note that the parameters marked underline indicate that these combinations will not reach the desired FPS. The table also indicates that Q-VR can accommodate a range of hardware, network and scene complexity conditions.

**System Energy Sensitivity.** As the predominant energy consumer on mobile systems, Fig.15 shows the normalized GPU energy efficiency of Q-VR under different hardware and network conditions. We estimate network module power by referring to the previous works [23, 25]. We also count the energy consumption of UCA and LIWC into the total energy consumption of Q-VR. Then we estimate the energy efficiency of Q-VR by normalizing its energy consumption to traditional local rendering in commercial VR device. The figure shows that Q-VR achieves an average of 73% energy reduction over the purely local rendering even though the collaborative rendering incurs network overhead. This is because the local mobile hardware in Q-VR only processes the most critical fovea area with high resolution instead of the entire frame. We also observe that in general increasing the network throughput can improve the energy efficiency of Q-VR. This is because Q-VR is able to achieve better performance under high network bandwidth while the power consumption of network model is typically less critical than that of the local GPU. Additionally, reducing GPU frequency will not always increase the energy benefit due to larger GPUs dynamic energy consumption.

## 7 DISCUSSION

**Eye-Tracking Performance and Accuracy** – In our work, we estimate the performance of eye-tracking system based on the publicly available stats from the state-of-the-art eye-trackers, which are implemented in HTC VIVE Pro Eye[53] and tobii Pio Neo 2 Eye[54]. The latest eye-tracker system is able to reach the refresh rate of 120 Hz and high accuracy of under 1 degree for detection. As we mentioned in Fig-2 in Sec-2.1, like motion sensors, state-of-the-art eye-trackers are operated in parallel with the graphics pipeline on their own frequencies [13, 20, 53]. Thus, the actual eye tracking latency is not in the critical path of the graphics pipeline as our focus. In the work. we count a sensor-data transmission latency (i.e., around 2 ms [13, 20]) in the end-to-end latency discussion. Due to the proprietary nature of the eye-tracking chip design, it is hard to estimate its standalone energy. Since it has been widely integrated in the current mobile VR SoCs such as Snapdragon[53], we believe its energy consumption is acceptable for modern VR applications.

**Design Choice of LIWC** – Since the dynamic fovea selecting is on the critical path of Q-VR pipeline, it requires low latency as well as online learning capability to quickly identify the balanced point for different realtime constraints. To make this design choice, we have investigated several research-based and commercial DNN accelerators [8, 19, 40, 61]. We found that some of them [40, 61] are too power hungry for mobile VR systems while the others, e.g., Google coral edge TPU [19] and Eyeriss[8], cannot provide the required performance. For example, Google coral edge TPUs need 10-20ms to process a DNN inference and the training process has to rely on high-end GPUs. To this end, we propose a lightweight realtime Q-Learning based approach, LIWC, which maps the user inputs to scene complexity using online updated lookup table. To match the design goals, we drastically simplify the fine-grained tuning space of the original Q-learning by indexing the motion information and eccentricity as limited delta tags to greatly reduce its latency, power and design complexity.

## 8 RELATED WORK

**Studies On Foveation Effects.** Since the foveation effect of human visual system can provide significant workload reduction without affecting user experience, it has been studied in various aspects such as foveated compression[58, 59] and foveated rendering [1, 20, 38, 43, 51, 60]. Several works[1, 20, 38] also conduct user survey on the user perception for foveated rendering. Our work follows their suggestion to constrain resolution manipulation for the periphery layers to guarantee user perception in Q-VR design.

Recent work also employs the foveation effect to reconstruct the low-resolution image for VR/AR display using neural networks[24, 28]. Comparing with them, our work cooperates with mobile hardware and network resource to improve the performance of the mobile VR system. By exploring the accelerator-level parallel, the expanded high-quality fovea area is rendered fast and promptly while the resolution of periphery area is reduced to save the overall data transmit.

**Collaborative Computing.** There have been several works [7, 21, 26, 30, 31, 64] that improve the system performance by allocating part of the workload to multiple accelerators. In the computing graphics domain, works [7, 31, 36] have either enabled caching mechanism to store all the pre-rendered scenes [7] or employed static collaborative techniques between dynamic objects and background [31]. We provide a lengthy discussion about their issues concerning complex modern VR applications in Section 2.1. In contrast, Q-VR provides desirable Quality of Experience (QoE) for a wide range of VR applications, hardware, and network conditions by effectively leveraging the computing capability of the increasingly powerful hardware of both mobile systems and cloud servers. In the general-purpose application domain, Neurosurgeon [26] profiles the computing latency and data size for DNN layers and uses the information to identify the best static partition point.

Gables[21] refines the roofline model to estimate the collaborative computing performance among multi-accelerator on Mobile SoC.

## 9 CONCLUSION

Looking into the future, the state-of-the-art mobile VR rendering strategies become increasingly difficult to satisfy the realtime constraints for processing high-quality VR applications. In this work, we provide a novel software-hardware co-design solution, named *Q-VR*, to enable future low-latency high-quality mobile VR systems. Specifically, the software-level design of Q-VR leverages human visual effects to translate a difficult global collaborative rendering problem into a workable scope while the hardware design enables a low-cost local-remote latency balancing mechanism and deeper pipeline optimizations. Evaluation results show that our Q-VR achieves an average end-to-end performance speedup of 2.2X (up to 3.1X) and a **4.1x** frame rate improvement over the state-of-the-art static collaborative VR designs.

## ACKNOWLEDGMENT

This is research is partially supported by University of Sydney faculty startup funding, Australia Research Council (ARC) Discovery Project DP210101984 and Facebook Faculty Award. This research is partially supported by U.S. DOE Office of Science, Office of Advanced Scientific Computing Research, under the CENATE project (award No. 66150), the Pacific Northwest National Laboratory is operated by Battelle for the U.S. Department of Energy under contract DEAC05-76RL01830.